\documentclass[a4paper]{jpconf}
\usepackage{graphicx}
\usepackage{subfigure}

\begin{document}
\title{Neutron spectroscopy with N$_2$-filled high-pressure large-volume spherical proportional counters}

\author{I Giomataris$^2$, S Green$^1$, I. Katsioulas$^1$, P Knights$^1$, I Manthos$^1$, \mbox{J Matthews$^1$,} T Neep$^1$, K Nikolopoulos$^1$, T Papaevangelou$^2$, \mbox{B Phoenix$^1$}, J Sanders$^1$ and R Ward$^1$}

\address{$^1$ University of Birmingham, School of Physics and Astronomy, B15 2TT, Birmingham, UK}
\address{$^2$ CEA Saclay, Universite Paris-Saclay, IRFU, F-91191, Gif-sur-Yvette, France}

\ead{i.manthos@bham.ac.uk}

\begin{abstract}
Precise in-situ measurements of the neutron flux in underground laboratories is crucial
for direct dark matter searches, as neutron induced backgrounds can mimic the typical dark matter signal.
The development of a novel neutron spectroscopy technique using Spherical Proportional Counters is investigated.
The detector is operated with nitrogen and is sensitive to both fast and thermal neutrons through the 
 $^{14}$N(n, $\alpha$)$^{11}$B and $^{14}$N(n, p)$^{14}$C reactions. This method holds potential to be a safe, inexpensive, effective, and reliable alternative to $^3$He-based detectors. 
Measurements of fast and thermal neutrons from 
an Am-Be source with a Spherical Proportional Counter operated at pressures up to 2\,bar at Birmingham are discussed.
\end{abstract}

\section{Introduction and detector concept}
The concept of neutron spectroscopy dates back to neutron discovery in 1932. Despite several attempts towards an efficient neutron spectroscopy system, such measurements remain cumbersome and detailed neutron spectra are sparse both in scientific laboratories and industrial sites. 
The most widely used approach to-date relies on $^3$He-based detectors, which are particularly expensive given the scarcity of $^3$He. Neutron spectroscopy using Spherical Proportional Counter (SPC) \cite{Giomataris:2008ap}, filled with a nitrogen-based mixture, may provide a viable alternative.
The detection principle~\cite{Bougamont:2015jzx}, exploits the following reactions:
\begin{eqnarray}
\nonumber ^{14}N + n &\rightarrow {}^{14}C + p + 625\,{\rm keV}, \\
\nonumber ^{14}N + n &\rightarrow {}^{11}B + \alpha - 159\,{\rm keV}
\end{eqnarray}
It exhibits all major advantages of $^3$He-based detection
at a small fraction of the cost.

\begin{figure}[h]
\centering
\subfigure[\label{fig:fig1a}]{\includegraphics[width=12pc]{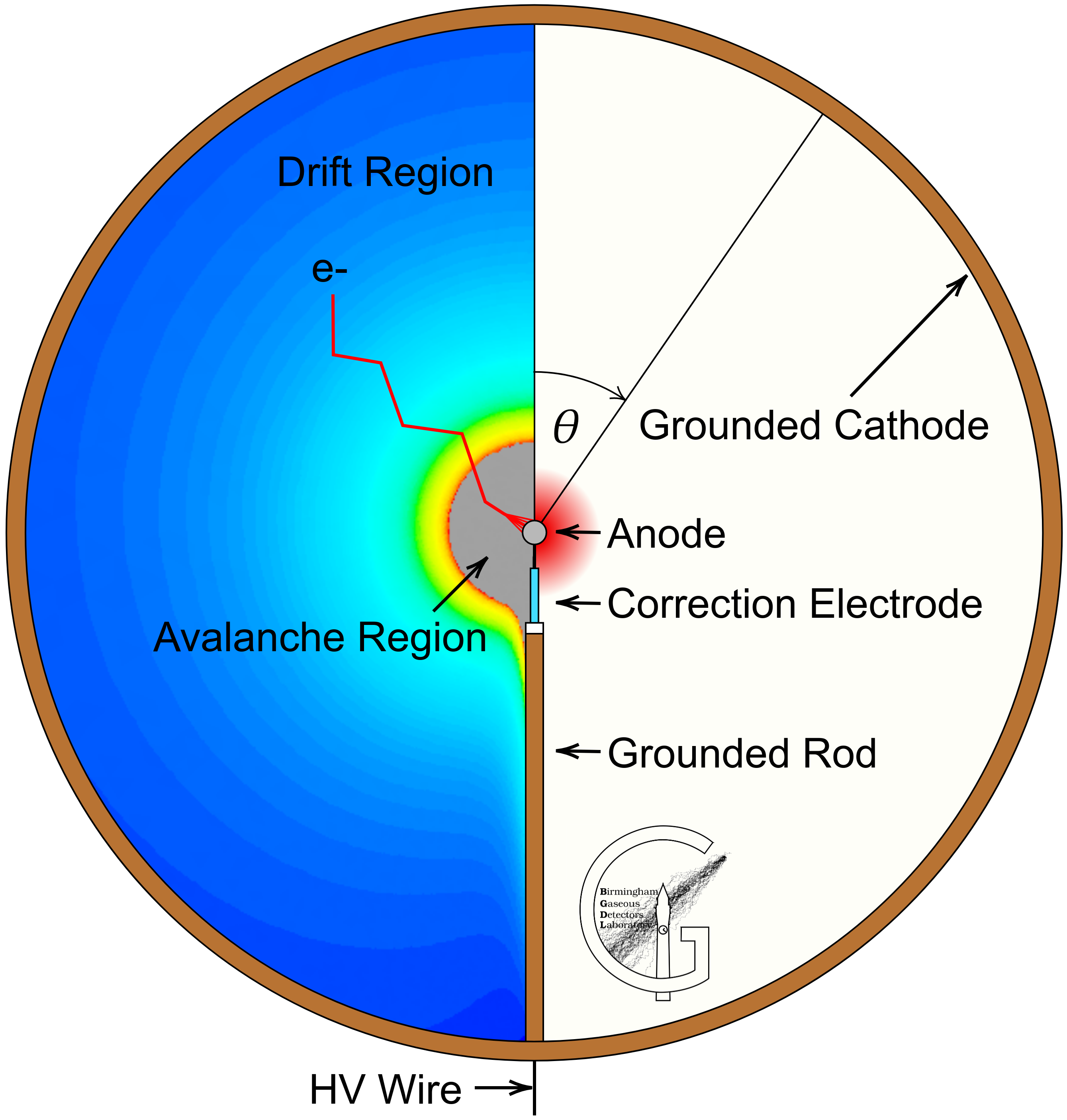}}
\subfigure[\label{fig:fig1b}]{\includegraphics[width=12pc]{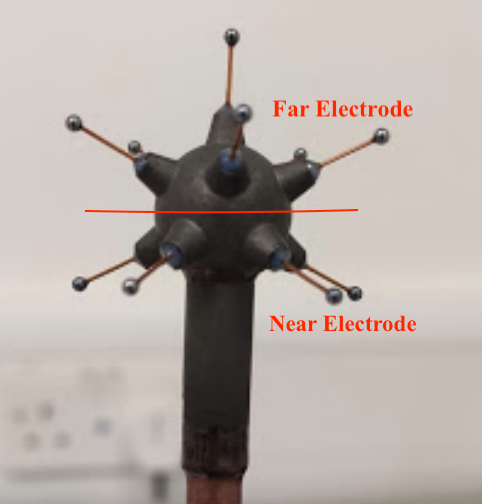}}
\caption{\subref{fig:fig1a} A schematic of the Spherical Proportional Counter and its principle of operation. \subref{fig:fig1b} A multi-anode read-out structure with 11-anodes.\label{fig1}}
\end{figure}

The SPC is a gaseous detector with a spherical geometry. The vessel is grounded and serves as the cathode. A sensor is placed at the centre of the vessel, which comprises one or more anodes. Through the anodes of the sensor, the high voltage is applied and the signal is read-out. The sensor is supported by a grounded metallic rod, as shown in Fig.~\ref{fig:fig1a}. 

The SPC exhibits low electronic noise even for large volumes. In the ideal case of a single anode floating at the detector centre, the capacitance is practically independent of the size:
\begin{equation}\label{eq:eq1}
C=4\pi \epsilon\epsilon_{0}\frac{r_{c}r_{\alpha}}{r_{c}-r_{\alpha}}\approx 4\pi \epsilon\epsilon_{0}r_{\alpha}
\end{equation}
where $r_{c}$ and $r_{a}$ are the cathode and anode radius, respectively.
The radial electric field magnitude is:
\begin{equation}\label{eq:eq2}
E(r)\approx \frac{V_{0}}{r^{2}}r_{a}
\end{equation}
where $V_{0}$ is the bias of the central anode. The electric field magnitude scales as $1/r^{2}$ with the distance $r$ from the centre, separating the volume to a drift and an amplification region.
For large detectors or detectors operating at high pressure, the small ratio of the electric field to the gas pressure increases the probability for electron attachment and recombination. 
A solution to this problem is the use of a multi-anode structure, called ACHINOS~\cite{Giganon:2017isb}, which decouples drift and amplification fields. This advancement improves the field homogeneity in the detector volume, provides efficient charge collection with high gain and allows increased target masses. Such a sensor~\cite{Giomataris:2020rna} is shown in Fig.~\ref{fig:fig1b}. 

\section{Neutron measurements with SPC}
\begin{figure}[h]
\centering
\subfigure[\label{fig2a}]{\includegraphics[width=18pc]{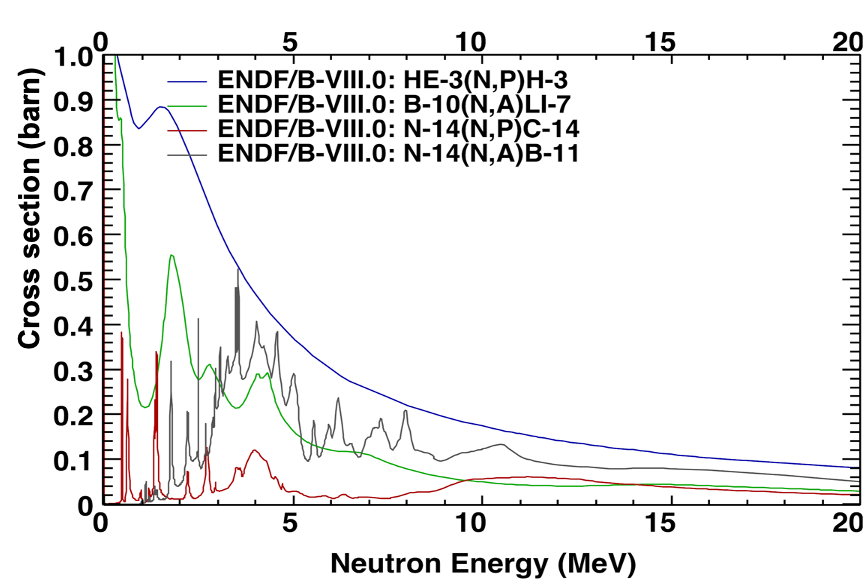}}
\subfigure[\label{fig2b}]{\includegraphics[width=16pc]{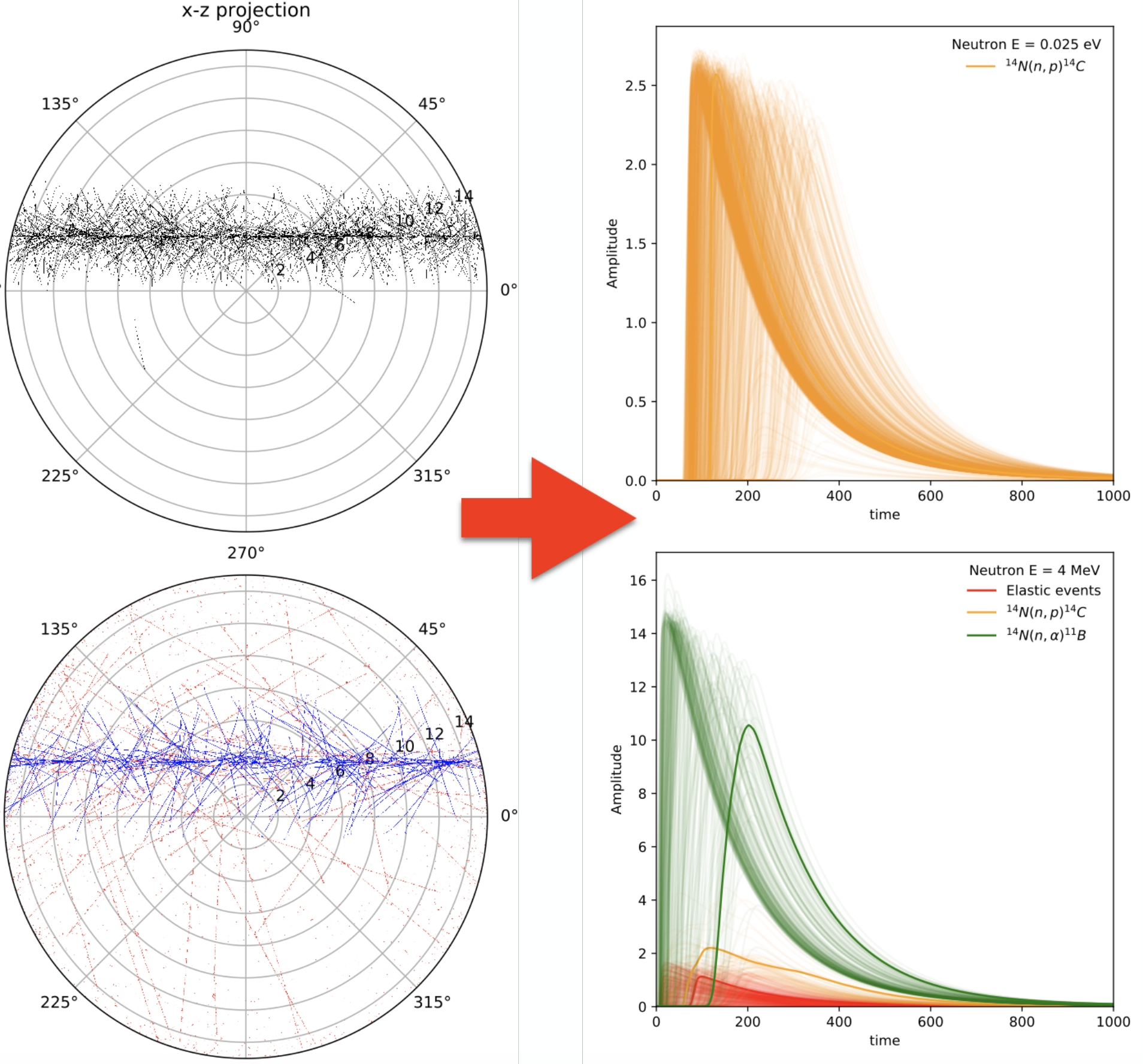}}
\caption{\subref{fig2a} Neutron absorption cross-section for different targets. \subref{fig2b} Simulated neutron events and the obtained pulses in the SPC for thermal (top) and fast neutrons (down).\label{fig2}}
\end{figure}

 The large SPC volume, the capability of operation at high pressure and the high interaction probability of neutrons with N$_2$, approaching this of $^3$He as shown in Fig.~\ref{fig2a}, makes the detector efficient for both fast and thermal neutron detection. Examples of simulated~\cite{Katsioulas:2020ycw} fast and thermal neutrons interactions in the SPC and the resulting pulses are shown in Fig.~\ref{fig2b}.
 Potential wall effect, due to  reaction products reaching the detector walls, is suppressed thanks to the increased pressure. 

A 30\,cm diameter SPC equipped with an 11-anode ACHINOS sensor with 1\,mm anode diameter was operated at the University of Birmingham (UoB) using N$_2$ at pressures up to 2\,bar. The sensor was readout in two channels, as shown in Fig.~\ref{fig:fig1b}: The 5 anodes closest to the supporting rod (``near” side), and the rest (``far” side).
\begin{figure}[h]
\centering
\subfigure[\label{fig3a}]{\includegraphics[width=10pc]{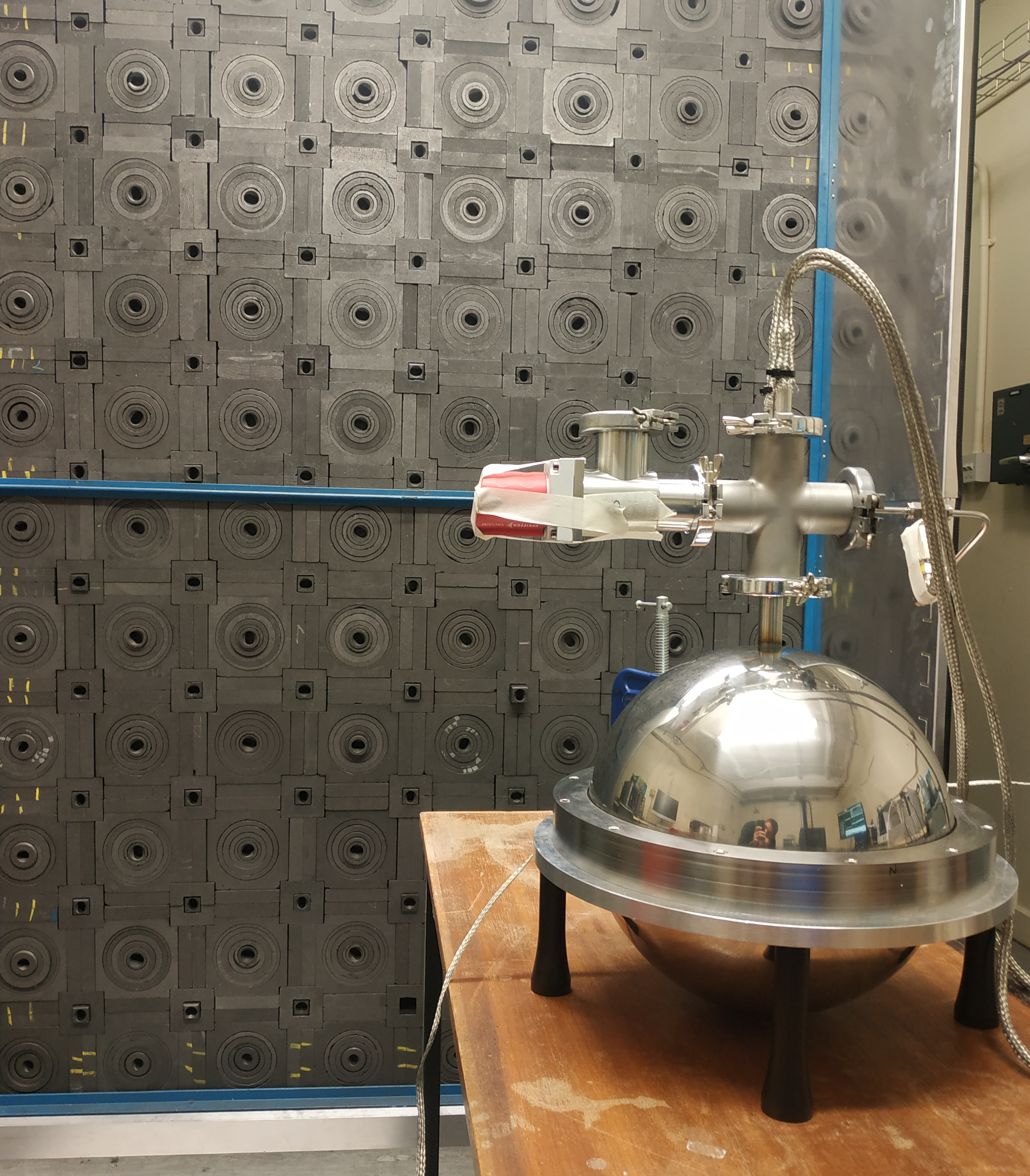}}
\subfigure[\label{fig3b}]{\includegraphics[width=16.5pc]{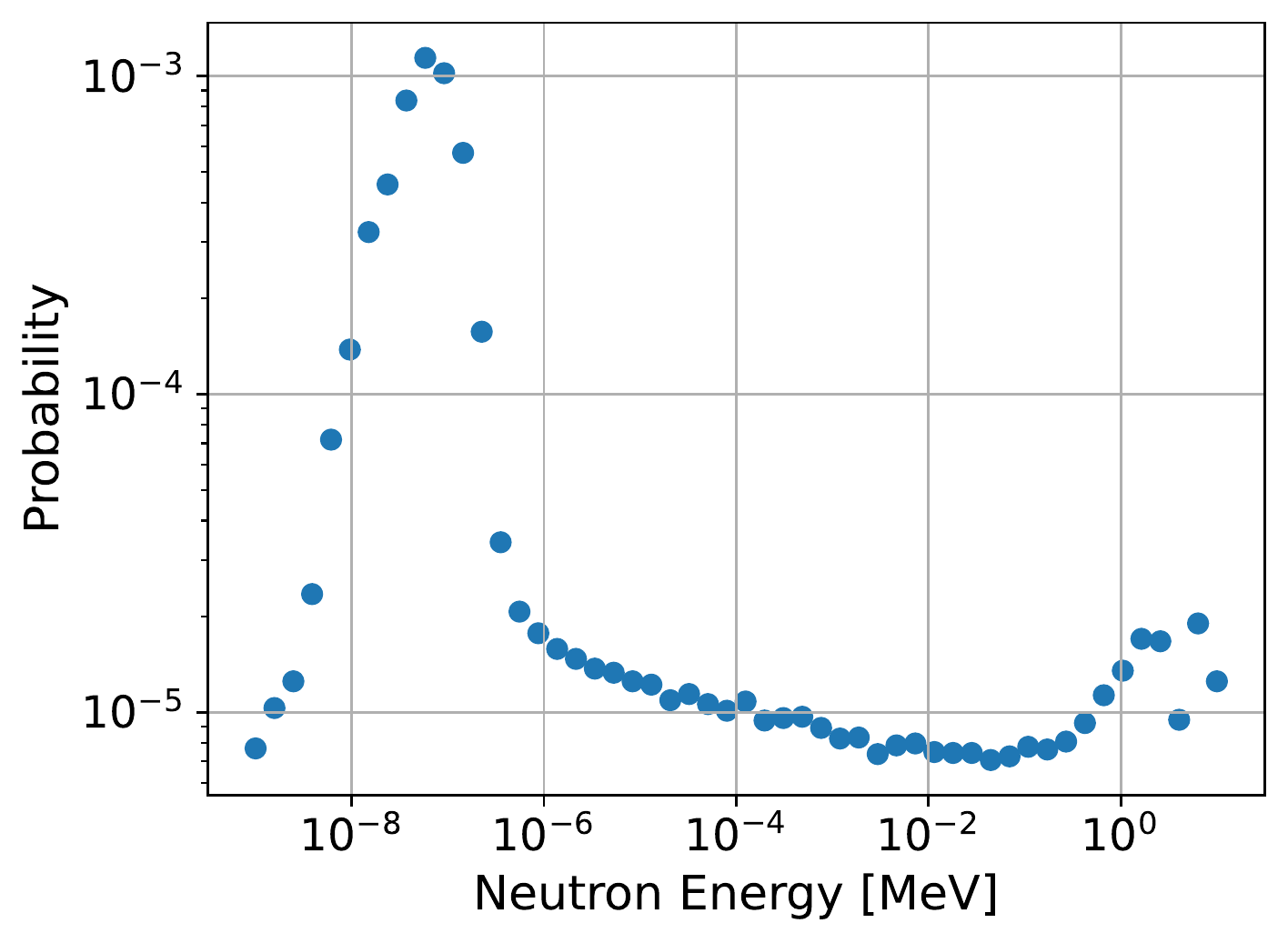}}
\caption{\subref{fig3a} The 30\,cm in diameter SPC in front of the graphite stack. \subref{fig3b} Probability for an emitted neutron to reach the detector, as a function of its energy at the detector.\label{fig3}}
\end{figure}
Neutrons are provided by an $^{241}$Am$^9$Be source (A = $2.6 \times 10^{6}$\,Bq) placed inside the graphite stack shown in Fig.~\ref{fig3a}. Emitted neutrons thermalise through interactions with the graphite. The probability of a neutron to reach the detector is estimated with simulations to be $5.22\times 10^{-3}$, while the energy distribution of the neutrons at the SPC is shown in Fig.~\ref{fig3b}.
 
 \section{Measurements results}
 In the following, SPC measurements of thermal and fast neutrons using N$_2$ at pressures up to 2\,bar are presented.
 Pulse shape quality criteria, such as pulse rise and fall time,  full-width at half-maximum (FWHM), and ratio of amplitude to FWHM, are applied. 
\begin{figure}[h]
\centering
\subfigure[\label{fig4a}]{\includegraphics[width=13pc]{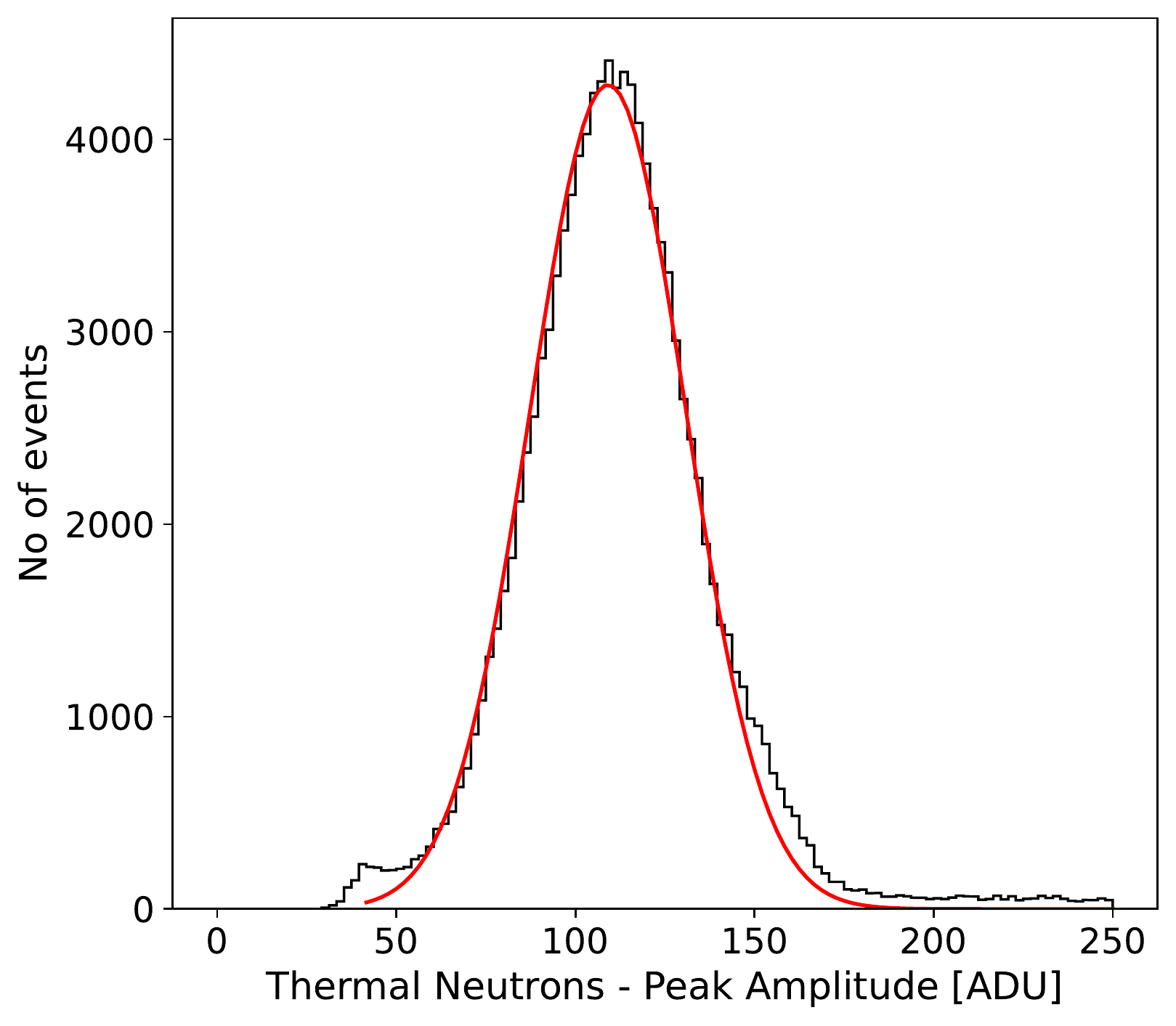}}
\subfigure[\label{fig4b}]{\includegraphics[width=13pc]{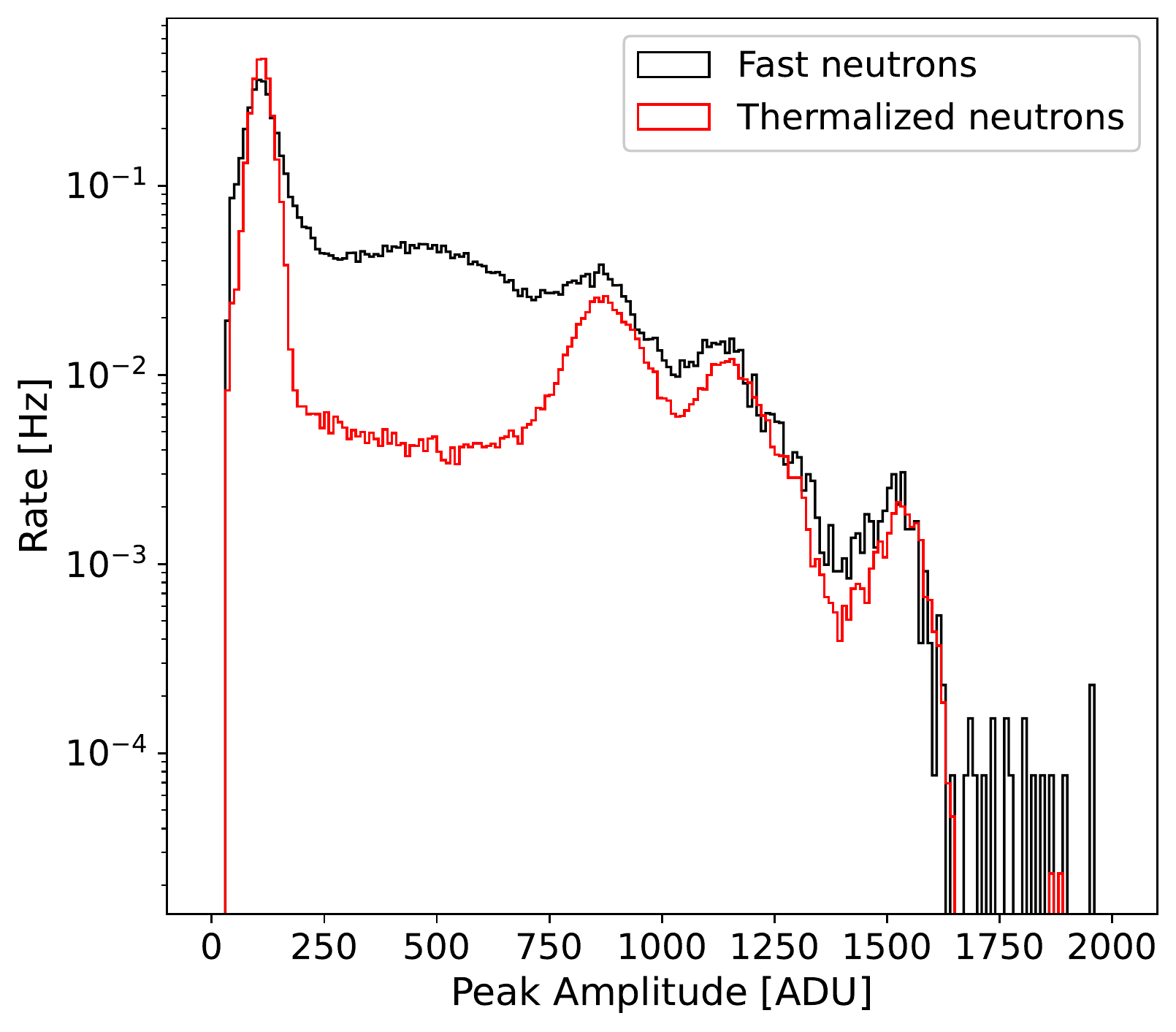}}
\caption{Data with N$_2$ at 1.5\,bar and 4.5\,kV anode voltage: \subref{fig4a} Thermal neutrons amplitude distribution; and \subref{fig4b} comparison of amplitude distribution for thermal and fast neutrons.\label{fig4}}
\end{figure}
The pulse amplitude obtained for the thermal neutrons in the near channel at 1.5\,bar with 4.5\,kV  anode voltage is presented in Fig. \ref{fig4a}. The peak at 109 ADU corresponds to the recoil energy of 625~keV and exhibits a resolution of approximately 20\%.
The observed rate of 5\,Hz corresponds in a detection efficiency of approximately $3.7 \times 10^{-4}$, in agreement with simulations. In Fig. \ref{fig4b}, the amplitude distributions obtained for fast and thermal neutrons are compared. The $\alpha$ particles from the $^{222}$Rn decay chain are also observed. The $^{222}$Rn peak 
at 868\,ADU confirms the thermal neutron peak position.
\begin{figure}[h]
\centering
\subfigure[\label{fig5a}]{\includegraphics[width=18pc]{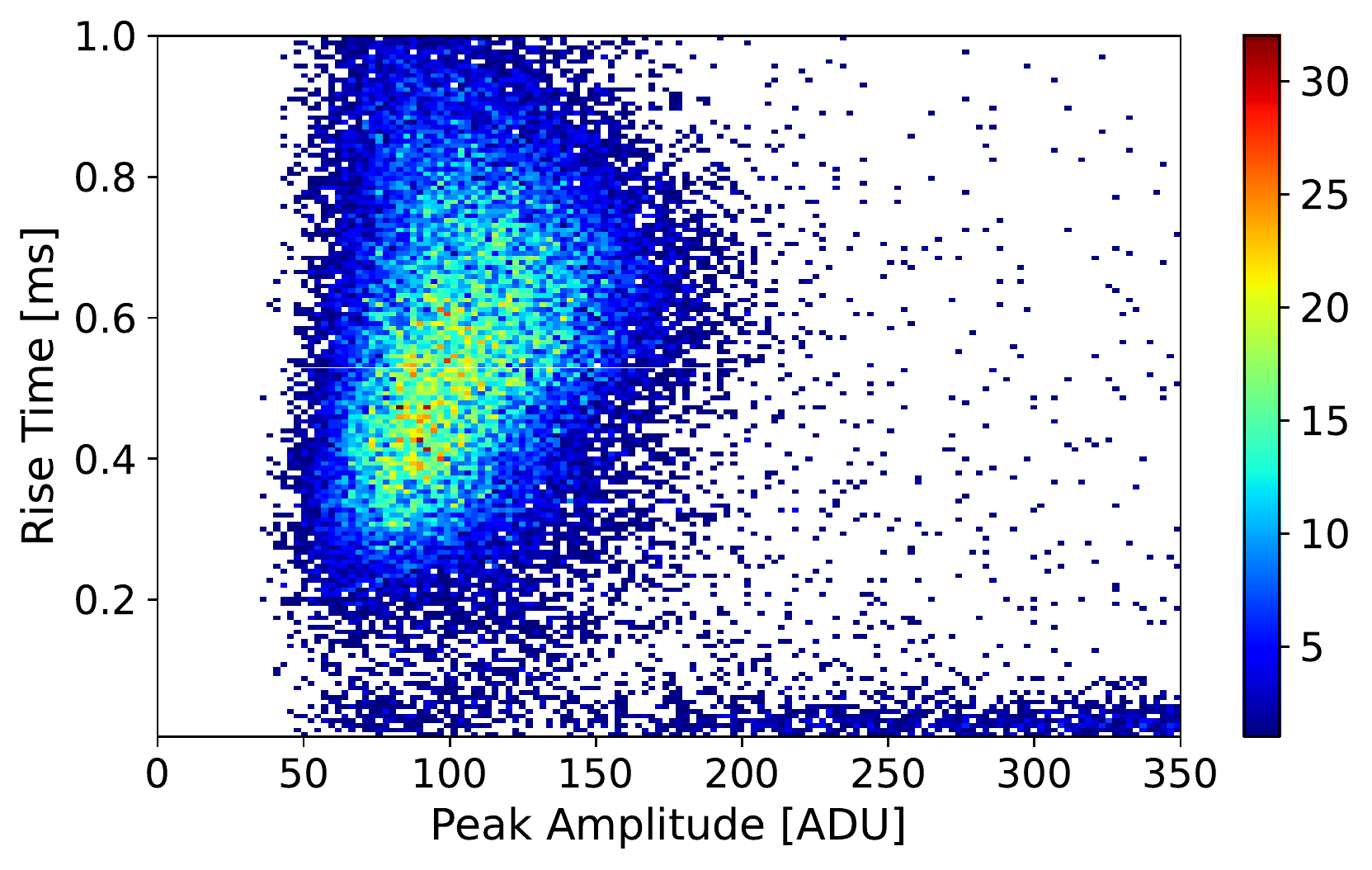}}
\subfigure[\label{fig5b}]{\includegraphics[width=13pc]{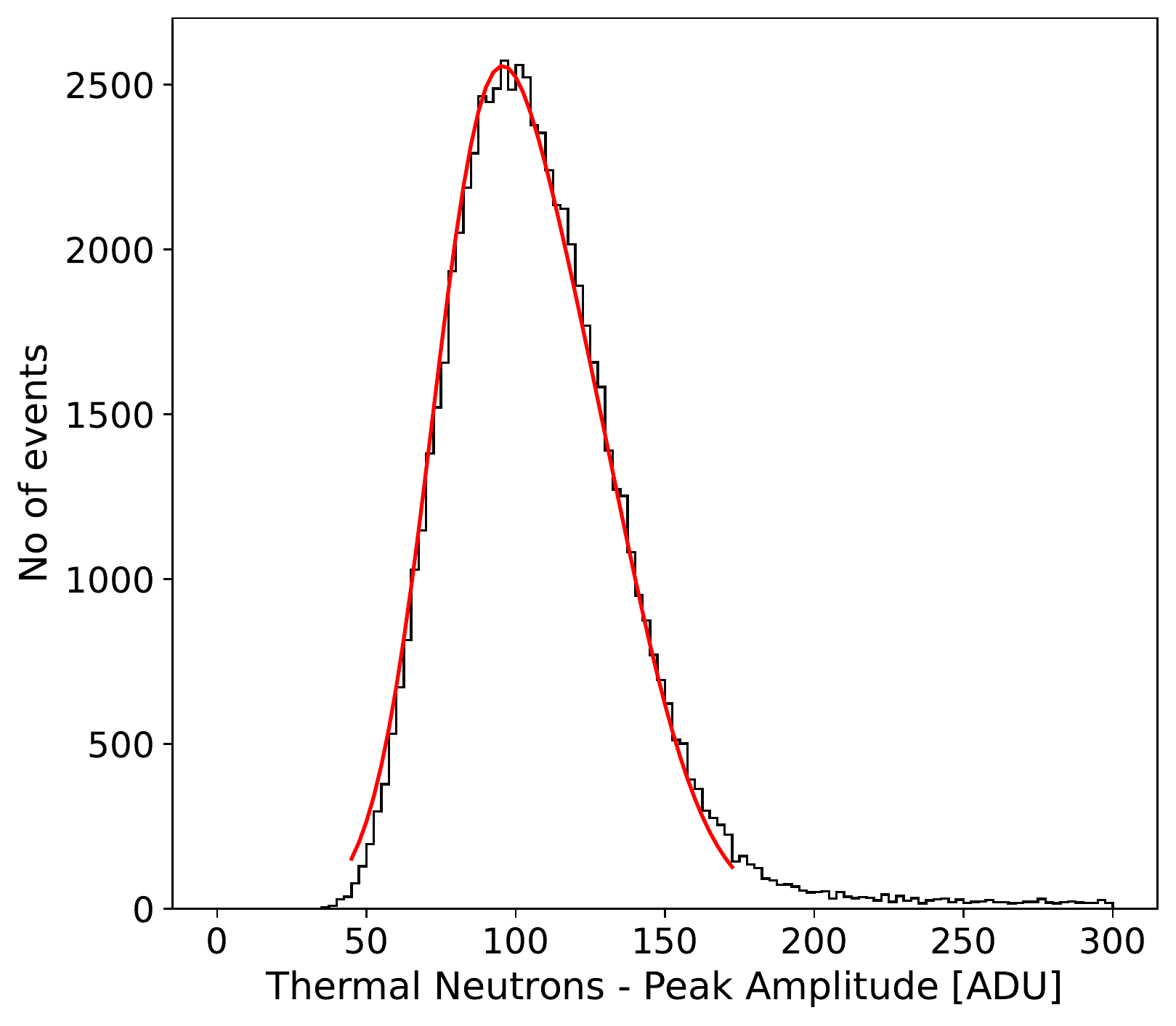}}
\caption{Data with N$_2$ at 2\,bar and 4.95\,kV anode voltage: \subref{fig5a} Rise time versus amplitude; and \subref{fig5b} Thermal neutrons amplitude distribution.\label{fig5}}
\end{figure}

Detection of thermal neutrons at operating pressures of 2\,bar is demonstrated in Fig.~\ref{fig5} at a rate of approximately  8\,Hz. The pulse rise time versus amplitude clearly discriminates events originated from thermal neutrons as shown in Fig.~\ref{fig5a}, while  in  Fig.~\ref{fig5b} the peak observed at 104\,ADU is estimated with a resolution of approximately 26\%.

\section{Conclusions}
SPCs filled with N$_2$ is a promising alternative to $^3$He-based detectors. Fast and thermal neutron detection with a N$_2$-filled SPC at high pressures of up to 2\,bar is demonstrated for the first time. Further detailed studies are planned, including at the Boulby Underground Laboratory.

\ack{This research is supported by European Union’s H2020 Research and Innovation Programme under Marie Sk\l{}odowska-Curie grants  neutronSphere (845168) and DarkSphere (841261).}

\section*{References}
\bibliographystyle{plain}
\bibliography{mybibliography}

\end{document}